\documentclass[12pt, preprint]{aastex}
\usepackage{graphicx}

\newcommand{\src}{KS\,1731$-$260}

\begin{document}

\title{The Mass and Radius of the Neutron Star in the Bulge 
Low-Mass X-ray Binary KS\,1731$-$260}

\author{Feryal \"Ozel$^1$, Andrew Gould$^2$, and Tolga G\"uver$^1$}

\affil{$^1$University of Arizona, Department of Astronomy, 933 N. Cherry 
Ave., Tucson, AZ 85721, USA; fozel@email.arizona.edu,
tguver@email.arizona.edu}
\affil{$^2$ Department of Astronomy, Ohio State University, 
140 W. 18th Ave., Columbus, OH 43210, USA;
gould@astronomy.ohio-state.edu}

\begin{abstract}
  Measurements of neutron star masses and radii are instrumental for
  determining the equation of state of their interiors, understanding
  the dividing line between neutron stars and black holes, and for
  obtaining accurate statistics of source populations in the
  Galaxy. We report here on the measurement of the mass and radius of
  the neutron star in the low-mass X-ray binary KS\,1731$-$260. The
  analysis of the spectroscopic data on multiple thermonuclear bursts
  yields well-constrained values for the apparent angular area and
  the Eddington flux of the source, both of which depend in a distinct
  way on the mass and radius of the neutron star. The binary
  KS\,1731$-$260 is in the direction of the Galactic bulge, allowing a
  distance estimate based on the density of stars in that
  direction. Making use of the Han \& Gould model, we determine the
  probability distribution over the distance to the source, which is
  peaked at 8~kpc. Combining these measurements, we place a strong
  upper bound on the radius of the neutron star, $R \le 12$~km, while
  confining its mass to $M \le 1.8$~M$_{\sun}$.

\end{abstract}


\section{Introduction}

A direct probe of the equation of state of cold, ultradense matter is
the measurement of the radii of neutron stars, whose cores are
comprised of the densest matter in the current universe.  Radii
measurements can distinguish between a variety of interior
compositions, as well as the strength of the many-body nuclear force
(Lindblom 1992; Lattimer \& Prakash 2001; Read et al.\ 2009;
\"Ozel \& Psaltis 2009). In addition, the mass distribution of 
neutron stars can give insights to the outcomes of supernova
explosions and evolutionary tracks in binaries. If sufficiently high,
a mass measurement can also lead by itself to constraints on the
composition and the interactions of the interior (Demorest et al.\
2010; \"Ozel et al. 2010a).

While precise mass measurements of neutron stars have been possible
for decades, significant progress in the determination of radii of
neutron stars has been made only recently. There are, by now, a
handful of neutron stars for which both a radius and a mass
measurement have been achieved. Some of these spectroscopic
observations have been carried out during the quiescent episodes of
accreting sources that reside in globular clusters (Rutledge et al.\
2001; Heinke et al.\ 2006; Webb \& Barret 2007; Guillot et al.\
2010). They have led to significant, albeit correlated, constraints in
the mass and radius of neutron stars.

Another way to measure the neutron star mass and radius is through a
combination of spectroscopic phenomena observed from their surfaces
during thermonuclear X-ray bursts (van Paradijs 1979; \"Ozel 2006).
The repeated high count-rate bursts allow a measure of the apparent
angular area of the neutron star over a wide range of temperatures
using time-resolved spectra. Furthermore, very luminous X-ray bursts
(called photospheric radius expansion, or PRE, bursts) cross the local
Eddington flux at the neutron star surface and allow us to obtain a
measure of the neutron star mass (as corrected for general
relativistic effects). Combined with the distance measurement to the
source, these quantities can be converted into independent
measurements of the neutron star mass and radius. This approach led to
the determination of the masses and radii of several neutron stars
(\"Ozel et al.\ 2009; G\"uver et al.\ 2010a, b) and enabled the
measurement of the pressure of neutron star matter above nuclear
saturation density (\"Ozel et al.\ 2010b).

In this paper, we place strong constraints on the mass and radius of
\src\ based on the analysis of thermonuclear bursts observed with the 
Rossi X-ray Timing Explorer.  The transient X-ray burster \src\ was
discovered in 1989 during observations with the TTM/Kvant telescope on
board the Mir station (Sunyaev et al.\ 1990). Three Type I bursts were
seen during the immediate follow-up observations, confirming the
presence of a neutron-star in this X-ray binary.  RXTE detected a
total of 24 Type-I bursts from \src, which allow a systematic study of
the spectra obtained during the bursts and the measurement of the
surface area and the Eddington limit of this neutron star (Galloway et
al.\ 2008; G\"uver et al.\ 2010c, d).

In order to convert the measurement of the apparent angular area and
Eddington flux from the surface of a neutron star into a measurement
of its mass and radius, we need an estimate of the distance to the
source. In earlier work, we used sources in globular clusters with
known distances (\"Ozel et al.\ 2009; G\"uver et al.\ 2010b) or
measured the source distance using the method of red clump stars
(G\"uver et al.\ 2010a). In this paper, we use the fact that \src\
lies within the Galactic bulge, which has a limited spatial extent in
the Galaxy. Even though the further localization of the source within
the bulge is not easy to achieve, it nevertheless allows us to place a
strong upper bound on the mass and radius of the neutron star.

In Section 2, we present the constraints on the distance to \src.  In
Section 3, we summarize the results from the analysis of the spectra
of thermonuclear bursts from this source and show the resulting
constraints on its mass and radius. In Section 4, we compare these
measurements to other existing mass-radius measurements.

\section{The Distance to \src}

\src\  lies at galactic coordinates $(l,b)=(1.06,3.65)$, i.e., almost 
exactly at the position of Baade's window (reflected in the Galactic
plane). Its position toward the bulge allows an estimate of its
distance based on the distribution of stars in that direction.  Having
no prior information on the distance to \src, we will assume that the
likelihood that it resides at a given distance $D$ is proportional to
the number density of stars along the line of sight to the bulge at
that distance.

For the stellar distribution along the line of sight we adopt the
model of \citet{hangould03}, which is based primarily on star counts,
and, without any adjustment, reproduces the microlensing optical depth
measured toward Baade's window.  For the bulge, \citet{hangould03}
normalize the ``G2'' K-band integrated-light-based bar model of
\citet{dwek95} using star counts toward
Baade's window from \citet{holtzman98} and \citet{zoccali00}.  For the
disk, they incorporate the model of \citet{zheng01}, which is a fit to
star counts. In this model, there is a 1~kpc hole at the center of
the Galactic disk, since this material would have gone into the
Galactic bulge according to the standard picture of pseudo-bulge
formation. We normalize the distance scale to a Galactocentric
distance of $R_0=8.0\pm0.4\,$kpc \citep{yelda10}. The position of the
source, being almost exactly at the position of Baade's window
(reflected in the Galactic plane), minimizes the extrapolation from
calibrating star counts to the adopted model.

\begin{figure*}
\centering
   \includegraphics[scale=0.75]{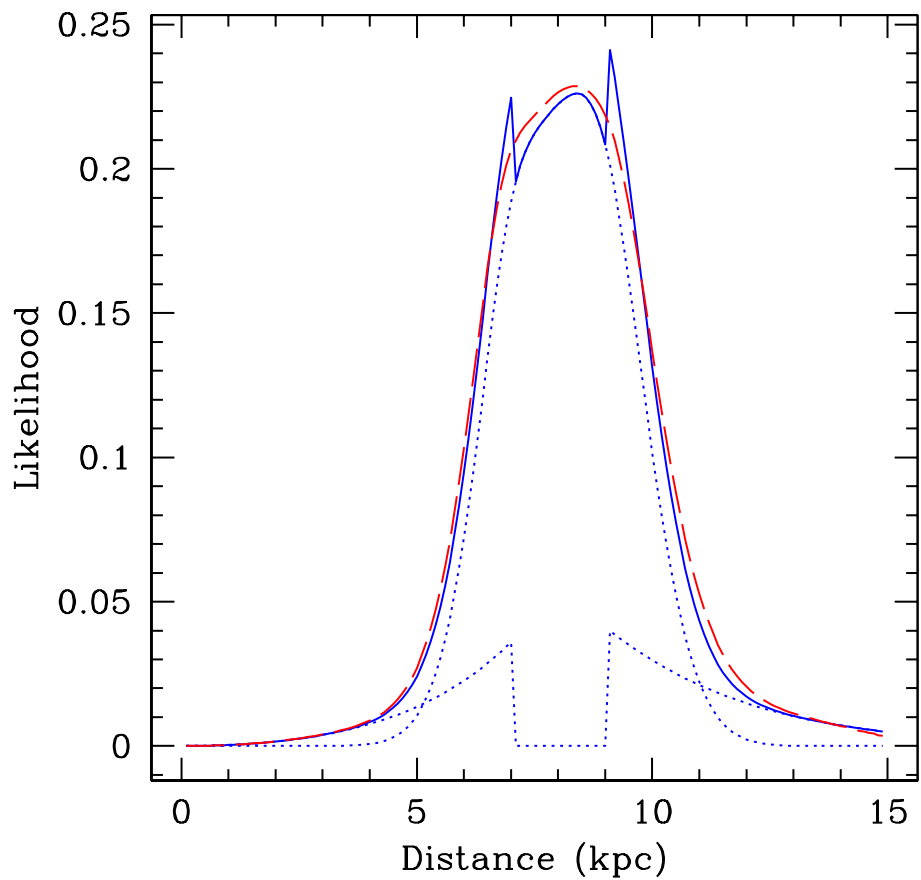} \caption{The solid line shows
   the likelihood of the distance to \src, which we have taken to be
   proportional to the number density of stars along the line of sight
   to the source, at each given distance, for an assumed distance to
   the Galactic center of 8.0~kpc. The dotted lines show the disk and
   bulge components of this distribution. The dashed line shows the
   total likelihood of the distance to \src\ when a Gaussian
   uncertainty of 0.4~kpc has been incorporated to the distance to the
   Galactic center. }
\label{fig:distance}
\end{figure*}

In Figure~1, we plot the stellar distribution along the line of sight
to \src, which we then take to be equal to the likelihood of the
source distance. The dotted lines show the disk and bulge
contributions separately, while the solid line shows the total
likelihood for a distance to the Galactic center of 8.0~kpc. The
dashed line shows the likelihood of the distance to \src\ when a
Gaussian uncertainty of 0.4~kpc has been incorporated to the distance
to the Galactic center. It is evident from the dashed curve that the
details of the prescription for the 1~kpc hole in the Galactic disk
are not important. The latter likelihood, which we use hereafter,
indicates that the source most probably lies at 5~kpc$<D<$11~kpc.

\section{Determination of the Neutron Star Mass and Radius from 
X-ray Burst Spectroscopy}

RXTE observed \src\ for a total of 483~ks until June 2007. We reported
in G\"uver et al.\ (2010c) the systematic analysis of the 24 bursts
detected in this time period. All of the details related to the
extraction of the spectra, detector response, background subtraction,
and the spectral fitting are discussed in G\"uver et al (2010c). We
only summarize here the results of the analysis of the time-resolved
continuum spectra in order to utilize them for the mass and radius
determination of the neutron star in this binary.

The 24 bursts we use here met our requirement that the persistent flux
prior to the burst was $<10\%$ of the peak burst flux, which minimizes
uncertainties related to background subtraction. We analyzed a total
of 1309 spectra from these bursts, out of which 1240 spectra (98\%)
gave acceptable values of reduced $\chi^2$. We measured the evolution
of the flux, the temperature, and the angular size using the spectra
obtained from 0.25, 0.5, or 1~s integrations during each burst. The
bursts follow highly reproducible tracks on the flux-temperature
diagram, as is shown in Figure~2. The slight decline in the
normalization at very low fluxes is consistent with the expected
increase of the color correction factor at low
temperatures. Incorporating this trend as a systematic uncertainty,
G\"uver et al.\ (2010c) reported an apparent angular area of $A = 88.4
\pm 5.1$~(km/10~kpc)$^2$, which corresponds to a radius of $9.4
\pm 0.3$~(km/10~kpc) during the cooling tails of bursts. The quoted
uncertainty in the radius includes statistical and systematic
uncertainties. The apparent angular area depends on the stellar
parameters according to
\begin{equation}
A=\frac{R^2}{D^2f_{\rm c}^4}\left(1-\frac{2GM}{Rc^2}\right)^{-1}\;,
\end{equation}
where $f_c$ is the color correction factor.

\begin{figure*}
\centering
   \includegraphics[scale=0.75]{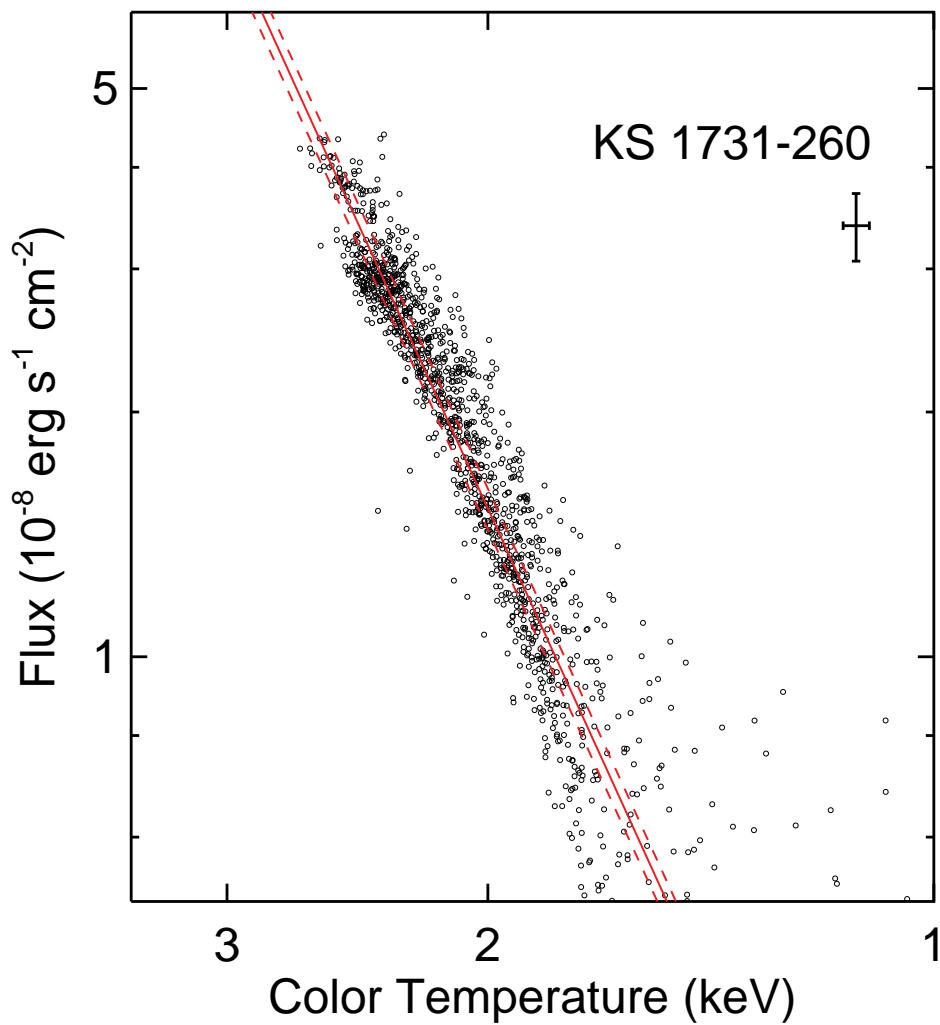}
   \caption{The evolution of the bolometric flux and the blackbody
     temperature during cooling tails of three thermonuclear X-ray
     bursts observed from \src. The bursts follow the $L
     \propto T^{4}$ relation, which is indicated by the solid line.}
\label{fig:hr}
\end{figure*}

\begin{figure*}
\centering
   \includegraphics[scale=0.75]{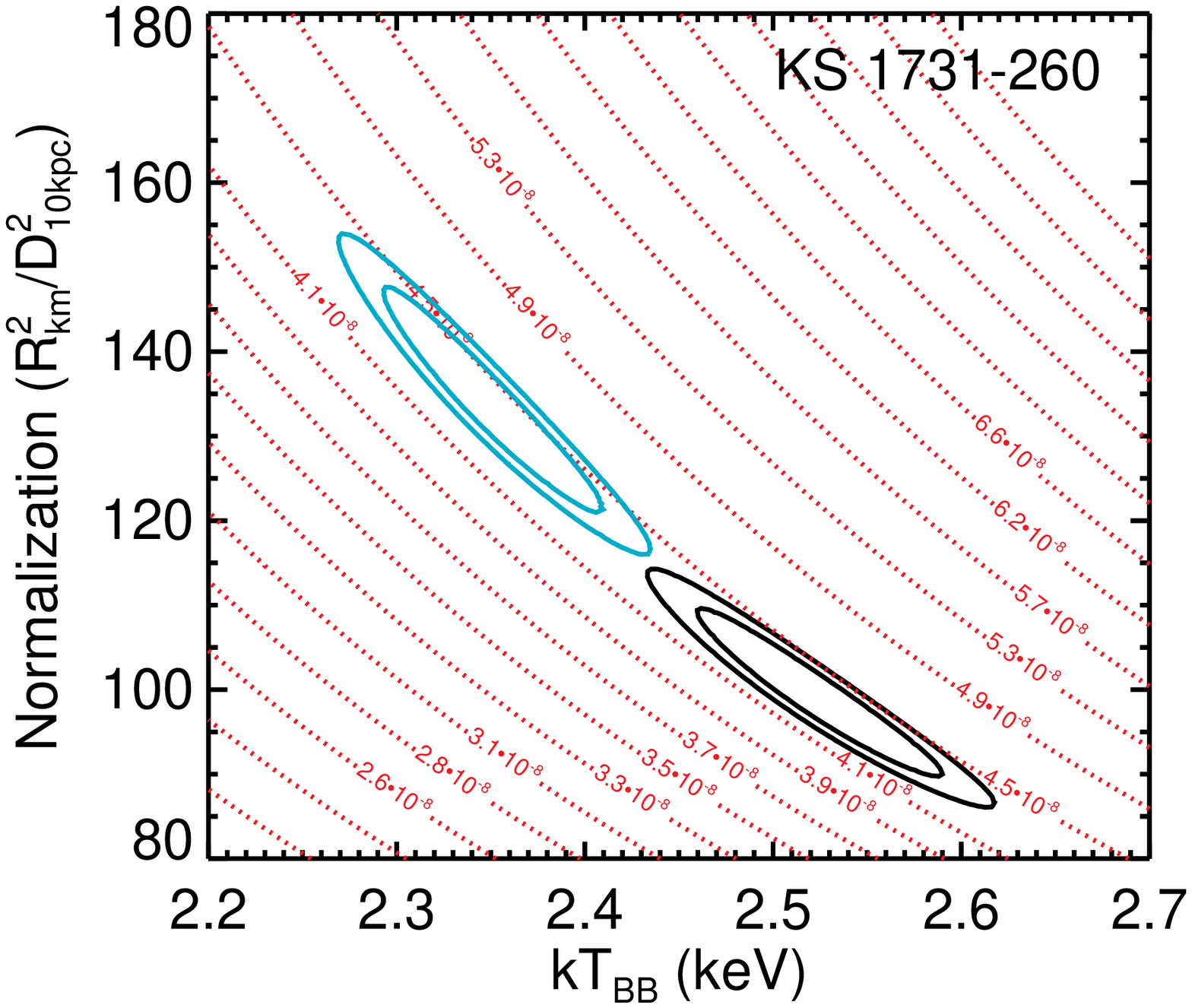} \caption{The confidence
   contours over the blackbody temperature and normalization during
   the touchdown moments of the two photospheric radius expansion
   bursts (in green and black, respectively) observed from \src.  The
   two bursts differ in both temperature and angular area, and they
   each have significant uncertainties in these variables, which are
   strongly correlated.  However, when the distribution is projected
   onto Eddington flux (red curves), the result is quite consistent
   between the two bursts and well constrained.}
\label{fig:edd}
\end{figure*}

Two out of the 24 bursts from \src\ showed clear evidence of
photospheric radius expansion, where the photosphere in highly
energetic burst events expands to many times the stellar radius, while
the flux and temperature follow a characteristic evolution (see
G\"uver et al. 2010d). We measured the Eddington flux in these events
at the touchdown moment when the photosphere returns to the stellar
surface, as indicated by a maximum in the temperature and a minimum in
the inferred area (as in \"Ozel et al.\ 2009 and G\"uver et al.\
2010d). For \src, we found an average touchdown flux of $(4.45 \pm
0.12)\times 10^{-8}$~erg~cm$^{-2}$~s$^{-1}$, where the quoted
uncertainty is purely formal. Because in sources with larger number of
photospheric radius expansion bursts we found systematic uncertainties
of the order of $\sim 5\%$ in the touchdown flux, we assign hereafter
a total uncertainty of $0.22 \times 10^{-8}$~erg~cm$^{-2}$~s$^{-1}$ in
order to be conservative. The touchdown flux depends on the stellar
parameters as
\begin{equation}
T_{\rm td}=\frac{GMc}{k_{\rm es}D^2}\left(1-\frac{2GM}{Rc^2}\right)^{1/2}, 
\end{equation}
where $G$ is the gravitational constant, $c$ is the speed of light,
and $k_{\rm es}$ is the opacity to electron scattering.

We assign independent probability distribution functions to the
distance $P(D)dD$, the touchdown flux $P(F_{TD})dF_{TD}$, and apparent
angular area normalization $P(A)dA$ based on the measurements and the
distance estimate discussed above. We take the probability
distributions over the touchdown flux and the apparent angular area to
be Gaussian. Converting the Eddington limit and the apparent angular
area to a mass-radius measurement requires a prior knowledge of the
hydrogen mass fraction $X$ in the atmosphere and a model for the color
correction factor $f_c$. In the absence of any information on the
composition of the accreted material for \src, we take a box-car
distribution for $X$ that covers the range 0$-$0.7. The color
correction factor, on the other hand, is obtained from modeling the
hot atmospheres of accreting, bursting neutron stars (e.g., Madej,
Joss, \& Rozanska\ 2004).  We take the box-car probability
distribution centered at $f_{c0}=1.35$ and a width $\Delta f_{c} =
0.05$ that is appropriate for a thermal flux in the range between
$\approx 1\% - 50\%$, which is seen in the cooling tails of bursts
(see G\"uver et al. 2010b, 2010c for a detailed discussion).

\begin{figure*}
\centering
   \includegraphics[scale=0.48]{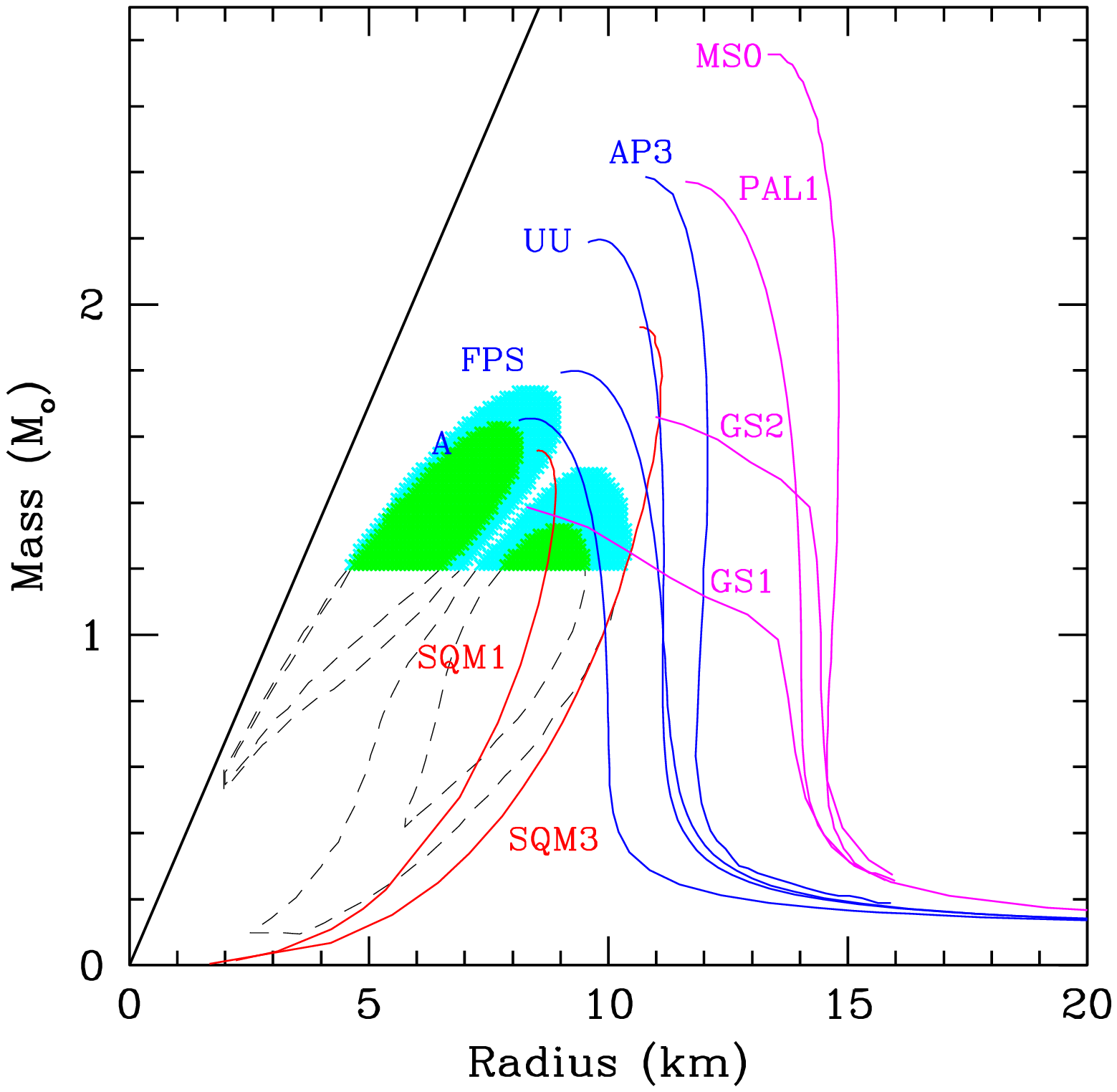}
   \includegraphics[scale=0.48]{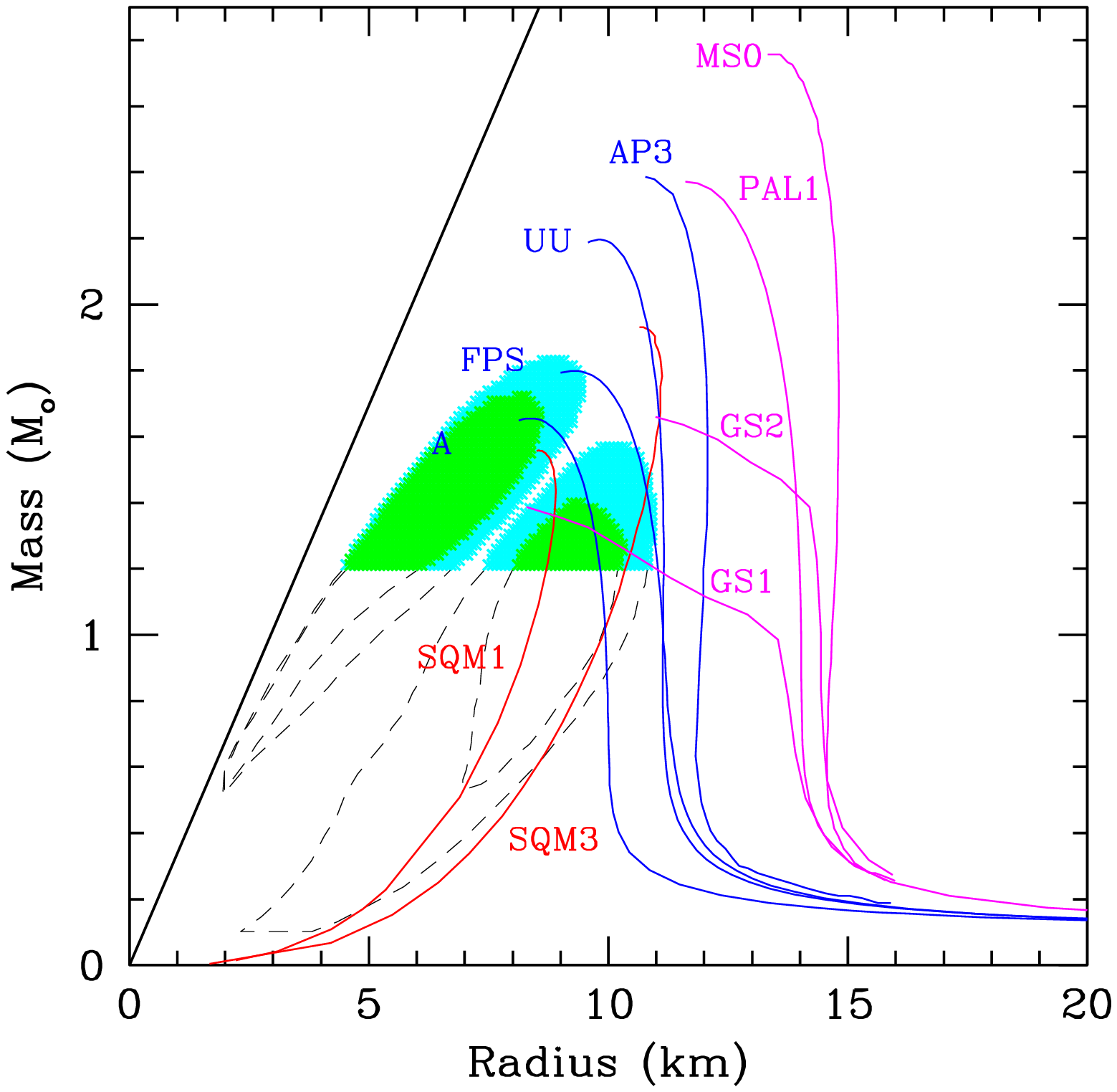} 
   \caption{{\it (Left)} The 68\% and 95\% confidence contours for the
   mass and radius of the neutron star in \src. The astrophysically
   relevant range of masses $M>1.2 M_\odot$ are shown as filled
   contours. The representative mass-radius relations for a select
   number of equations of state are also shown and include
   multi-nucleonic ones (A, FPS, UU, AP3), equations of state with
   condensates (GS1-2), strange stars (SQM1, SQM3), and meson-exchange
   models (MS0). The black line indicates the black hole event
   horizon. The descriptions of the various equations of state and the
   corresponding labels can be found in Lattimer \& Prakash (2001) and
   Cook, Shapiro \& Teukolsky (1994). {\it (Right)} The 68\% and 95\%
   confidence levels over neutron-star mass and radius calculated by
   using the maximum combined likelihood along lines of constant
   distance and hydrogen mass fraction rather than by integrating over
   these two nuisance parameters. The fact that these confidence
   levels are very similar to those shown in the left panel
   demonstrates that marginalization does not bias the results.}
\label{fig:massradius}
\end{figure*}

We calculate the probability distribution over the neutron star mass
and radius as in G\"uver et al.\ (2010a). Note that we have corrected
here an error in the Equation~(5) of \"Ozel et al.\ (2009) where the
term $7 GM/Rc^2$ should read $8 GM/Rc^2$. Figure~\ref{fig:massradius}
(left panel) shows the 68\% and 95\% confidence levels for the mass
and the radius of the neutron star in \src. Although the observed
apparent angular area and the touchdown flux are consistent with a
range of masses and radii for the neutron star, they nevertheless
provide a strong upper bound of $R<12$~km and $M<1.8 M_\odot$. In
Figure~\ref{fig:massradius}, we also identify the astrophysically
relevant range of neutron star masses $M>1.2 M_\odot$ and show the
combined constraints as filled contours. The allowed range of neutron
star radii is inconsistent with equations of state that predict large
neutron star radii, $\sim 12-15$~km. This is in agreement with the
earlier spectroscopic measurements of radii in bursting neutron stars
(\"Ozel et al. 2009; G\"uver et al.\ 2010a, b) as well as the
quiescent neutron stars in globular clusters (Webb \& Barret 2007;
Guillot et al.\ 2010).

When performing parameter estimation in models within a Bayesian
framework, two concerns often arise. First, there is no universally
accepted measure of goodness-of-fit, which makes it difficult to
assess whether data sets that are used to estimate the parameters are
consistent with the model. Second, in models with more than two
parameters, marginalizing (integrating) over the so-called nuisance
parameters in Bayesian analysis in order to infer the parameters of
interest can introduce biases. This is especially a concern when
measurement uncertainties in different parameters differ widely from
each other. In our case, the allowed logarithmic range of values over
distance is significantly larger than the uncertainties over the
touchdown flux and the apparent radius. These concerns have led
Steiner et al.\ (2010) to explore a different, ad hoc, interpretation
of the Eddington flux in their inference of the neutron star masses
and radii.

\begin{figure*}
\centering
   \includegraphics[scale=0.75]{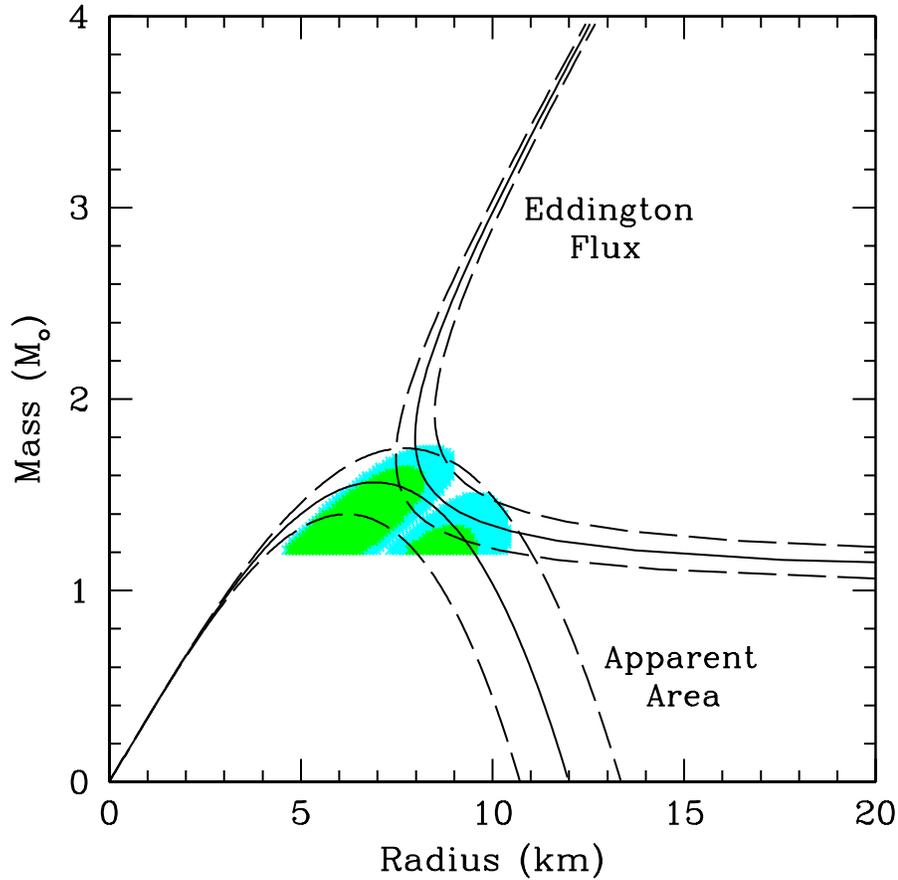} \caption{The contours in the
   mass-radius plane that correspond to the measurement of the
   Eddington flux and the apparent radius for a representative
   distance of 7~kpc. The dashed lines reflect the 1$\sigma$ errors on
   the measured quantities and model parameters. The fact that the two
   sets of contours intersect at the location of the confidence limits
   shown in the left panel of Figure~\ref{fig:massradius} and
   overplotted here (filled contours) demonstrate the consistency of
   the model with the data. }
\label{fig:lines}
\end{figure*}

In Figure~\ref{fig:lines}, we show the contours in the mass-radius
plane that correspond to each individual spectroscopic measurement,
taking into account the 1$\sigma$ ranges in the measured quantities
but fixing the distance to a representative value of 7~kpc, which is
within the region of highest likelihood (see Fig.~\ref{fig:distance})
and in the range of distances most favored by the data. The two sets
of contours intersect with each other within their 1$\sigma$ levels
and overlap with the combined confidence contours over mass and
radius. Even though this is not a quantitative measure of
goodness-of-fit, it shows the self-consistency of the theoretical
framework with the data.

In order to address the second concern about the effect of
marginalizing over distance, we performed the following test.  We
calculated the likelihood over mass and radius by using the {\it
maximum} of the combined likelihood along lines of constant distance
and constant hydrogen mass fraction $X$, instead of integrating over
these two parameters.  We show in the right panel of
Figure~\ref{fig:massradius} the resulting confidence contours over
mass and radius. The result is very similar to that shown in the left
panel of Figure~\ref{fig:massradius}, demonstrating that the
marginalization over the two most uncertain parameters, namely
distance and the hydrogen mass fraction, does not bias the mass and
radius measurement.

\section{Conclusions}

We used the spectroscopic measurements of the apparent angular area
and the Eddington flux during thermonuclear X-ray bursts from \src\ to
measure its mass and radius. We place a strong upper bound on the
radius of $R < 12$~km and confine the mass to $M< 1.8 M_\odot$. This
measurement challenges equations of state that predict radii larger
than $\sim$12~km.

We also explored whether, within the Bayesian parameter estimation
approach, 1. the theoretical framework and the data sets used here are
self-consistent and 2. the integration over the nuisance parameters
introduces biases in the confidence limits over the neutron star mass
and radius. We found that for a highly likely value of the distance,
the contours in the mass-radius plane that correspond to the apparent
radius and the Eddington limit are consistent with each other and with
the confidence contours over mass and radius within
1$\sigma$. Moreover, we showed that considering the most likely
configuration for each value of the distance and the hydrogen mass
fraction, as opposed to integrating over these two parameters, leaves
the results largely unaffected. These results justify our
identification of the touchdown point for the measurement of the
Eddington flux and our conclusion that the apparent angular area
during the cooling tail of the burst corresponds to the entire neutron
star surface.

\acknowledgments  
This work was supported in part by NASA ADAP grant NNX10AE89G and
Chandra Theory grant TMO-11003X. AG was supported by NSF grant
AST-0757888. We thank Andrew Steiner for pointing out an error in a
numerical factor in our earlier calculation of the Jacobian
transformation. We thank Dimitrios Psaltis for a careful reading of
the manuscript.


\begin{thebibliography}{37}

\bibitem[Cook et  al. (1994)]{1994ApJ...424..823C} Cook, G.~B., Shapiro, 
S.~L., Teukolsky, S.~A. 1994, \apj, 424, 823

\bibitem[Demorest et al.(2010)]{2010Natur.467.1081D} Demorest, P.~B., 
Pennucci, T., Ransom, S.~M., Roberts, M.~S.~E., \& Hessels, J.~W.~T.\
2010, \nat, 467, 1081

\bibitem[Dwek et al.(1995)]{dwek95} Dwek, E., et al. 1995,
\apj, 445, 716

\bibitem[Galloway et al. (2008)]{ga2008} Galloway, D.K., Muno M. P.,
  Hartman J. M., Psaltis D., Chakrabarty D.\ 2008, \apjs, 179, 360

\bibitem[Guillot et al.(2010)]{grb10} Guillot, S., Rutledge, 
R.~E., \& Brown, E.~F.\ 2010, arXiv:1007.2415

\bibitem[G{\"u}ver et al.(2010a)]{guveretal10a} G\"uver, T., 
\"Ozel, F., Cabrera-Lavers, A., \& Wroblewski, P.\ 2010a, \apj, 712,
964

\bibitem[G{\"u}ver et al.(2010b)]{guveretal10b} G\"uver, T., 
Wroblewski, P., Camarota, L., \"Ozel, F.\ 2010b, \apj, 719, 1807

\bibitem[Guver et al. (2010c)] {guveretal10c} G\"uver, T., Psaltis, D., \& 
\"Ozel, F.\ 2010c, \apj, submitted

\bibitem[Guver et al. (2010c)] {guveretal10d} G\"uver, T., \"Ozel, F., \& 
Psaltis, D. \ 2010d, \apj, submitted

\bibitem[Han \& Gould (2003)]{hangould03} Han, C. \& Gould, A. 
2003, \apj, 592, 172

\bibitem[Heinke et al. (2006)]{heinke} Heinke, C.~O., Rybicki, 
G.~B., Narayan, R., \& Grindlay, J.~E.\ 2006, \apj, 644, 1090

\bibitem[Holtzman et al.(1998)]{holtzman98} Holtzman, J.A.,
Watson, A.M., Baum, W.A., Grillmair, C.J., Groth, E.J., Light, R.M.,
Lynds, R., \& O'Neil, E.J., Jr.\ 1998, \aj, 115, 1946

\bibitem[Lattimer \& Prakash (2001)]{l2001} Lattimer, J.M. \& Prakash
  M., 2001, \apj, 550, 426

\bibitem[Madej et al. (2004)]{2004ApJ...602..904M} Madej, J., Joss, P.~C., 
\& R{\'o}{\.z}a{\'n}ska, A.\ 2004, \apj, 602, 904 

\bibitem[\"Ozel (2006)]{o2006} \"Ozel,~F. 2006, Nature, 441, 1115

\bibitem[\"Ozel et al. (2009)]{o2009}  {\"O}zel F., G{\"u}ver T., \&
Psaltis D., 2009, \apj, 693, 1775

\bibitem[\"Ozel \& Psaltis (2009)]{op09} {\"O}zel, F. \& Psaltis, 
D.\ 2009, \prd, 80, 103003

\bibitem[{\"O}zel et al.(2010)]{2010ApJ...724L.199O} {\"O}zel, F., 
Psaltis, D., Ransom, S., Demorest, P., \& Alford, M.\ 2010, \apjl,
724, L199

\bibitem[Rutledge et al. (2002)]{r2002} Rutledge, R.~E., Bildsten, L., 
Brown, E.~F., Pavlov, G.~G., Zavlin, V.~E., \& Ushomirsky, G.\ 2002,
\apj, 580, 413

\bibitem[Steiner et al.(2010)]{steiner10} Steiner, A.~W., 
Lattimer, J.~M., \& Brown, E.~F.\ 2010, \apj, 722, 33

\bibitem[van Paradijs (1979)]{v1979} van Paradijs, J. 1979, ApJ, 234, 609

\bibitem[Webb \& Barret (2007)]{webb} Webb, N.~A. \& Barret, D.\ 
2007, \apj, 671, 727

\bibitem[Yelda et al.(2010)]{yelda10} Yelda, S., Ghez, A.M., Lu, J.R.,
Do, T., Clarkson, W., \& Matthews, K. 2010, The Galactic Center: A
Window on the Nuclear Environment of Disk Galaxies, eds. M. Morris,
D. Q. Wang and F. Yuan, in press

\bibitem[Zheng et al.(2001)]{zheng01} Zheng, Z., Flynn, C.,
Gould, A., Bahcall, J.N., \& Salim, S.\ 2001, \apj, 555, 393

\bibitem[Zoccali et al.(2000)]{zoccali00} Zoccali, M., Cassisi,
S., Frogel, J.A., Gould, A., Ortolani, S., Renzini, A., Rich, R.M.,
\& Stephens, A.W.\ 2000, \apj, 530, 418

\end{thebibliography}
\end{document}